\documentclass[12pt,a4paper]{article}
\usepackage[latin1]{inputenc}
\usepackage{amsmath}
\usepackage{amsfonts}
\usepackage{textcomp}
\usepackage{amstext}
\usepackage{parskip}
\usepackage{a4wide}
\usepackage{graphicx}
\usepackage{url}                
\usepackage[small,it]{caption}
\usepackage{amssymb}
\usepackage{enumerate}
\usepackage{color}
\usepackage{url}
\let\origthanks\thanks
\renewcommand\thanks[1]{\begingroup\let\rlap\relax\origthanks{#1}\endgroup}

\usepackage{subcaption} 

\newcommand{\beq}{\begin{equation}}
\newcommand{\eeq}{\end{equation}}
\newcommand{\bea}{\begin{eqnarray}}
\newcommand{\eea}{\end{eqnarray}}

\newcommand{\Mbh}{M_{\rm BH}}
\newcommand{\Mpl}{M_{\rm P}}
\newcommand{\Lpl}{\ell_{\rm P}}

\begin{document}
\title{\textbf{Self-Completeness and the Generalized Uncertainty Principle}}
%
%
\author{Maximiliano Isi $^{a,}$\footnote{\textit{E-mail:} \texttt{misi@lion.lmu.edu}},  Jonas Mureika $^{a,}$\footnote{\textit{E-mail:} \texttt{jmureika@lmu.edu}}  and Piero Nicolini $^{b, c,}$\footnote{\textit{E-mail:} \texttt{nicolini@fias.uni-frankfurt.de}} \bigskip\\
 \textit{$^a$Department of Physics, Loyola Marymount University, } \\
 \textit{ Los Angeles, CA  90045-2659}\medskip\\
 \textit{$^b$Frankfurt Institute for Advanced Studies, Ruth-Moufang-Strasse 1}\\
 \textit{60438 Frankfurt am Main, Germany}\medskip\\
 \textit{$^c$Institut f\"{u}r Theoretische Physik, J. W. Goethe-Universit\"{a}t, }\\
\textit{Max-von-Laue-Strasse 1, 60438 Frankfurt am Main, Germany}
}

%

\maketitle

\begin{abstract}
\noindent
The generalized uncertainty principle discloses a self-complete  characteristic of gravity, namely the possibility of masking any curvature singularity behind an event horizon as a result of matter compression at the Planck scale.  In this paper we extend the above reasoning in order to overcome some current limitations  to the framework, including the absence of a consistent metric describing such Planck-scale black holes.  We implement a minimum-size black hole in terms of the extremal configuration of a neutral non-rotating metric,  which we derived by mimicking the effects of the generalized uncertainty principle via a short scale modified version of Einstein gravity.  In such a way, we find a self-consistent scenario that reconciles the self-complete character of gravity and the generalized uncertainty principle.
\end{abstract}

\section{Introduction}
It is a foregone conclusion that our classical understanding of gravitation is not applicable in the quantum regime.   A number of resolutions to this inadequacy involving modifications to spacetime structure have been proposed, including string inspired models and spin-loop networks.   A noted feature that has gained much traction over the last decade is the necessity of a minimal length scale that sets the quantum gravity threshold.  This provides a natural platform for self-regularization of quantum field theories \cite{kempf}, and furthermore allows for quantum gravity to be realizable in $(3+1)$-dimensions. 

Along these lines, it has been shown \cite{dvali1,dvali2,dvali3,euroantonio} that gravity may be considered \emph{self-complete}, in the sense that there exists a minimum horizon scale hiding  curvature singularities.  Specifically, this distance is defined by the confluence of the classical Schwarzschild radius $r_{\rm H}$ and the Compton wavelength $\lambda_{\rm C}$,
\beq
r_{\rm H} = \lambda_{\rm C} ~~~\Longrightarrow~~~\frac{2G\Mbh}{c^2} = \frac{h}{c\Mbh}~.
\eeq
This gives the mass of the lightest black hole
\beq
\Mbh \geq \sqrt{\frac{h c}{2G}} =\sqrt{\pi} \Mpl
\label{eq:selfcomplete}
\eeq
and, at the same time, the mass of heaviest quantum mechanical particle.
As a result the Planck scale $\Mpl= \sqrt{\hbar c/G}$ corresponds to the energy at which matter undergoes a transition from a particle phase to a black hole one.   By looking at the corresponding length scale, one learns that the Planck length $\Lpl\equiv\Mpl^{-1}$ is the minimal size for both  particles and black holes, which makes $\Lpl$ the smallest resolvable scale. 
From this perspective, the sub-Planckian world is dominated by light objects described by quantum mechanics, while the trans-Planckian world is dominated by classical objects described by GR.

\begin{figure*}
\centering
  \includegraphics[scale=1]{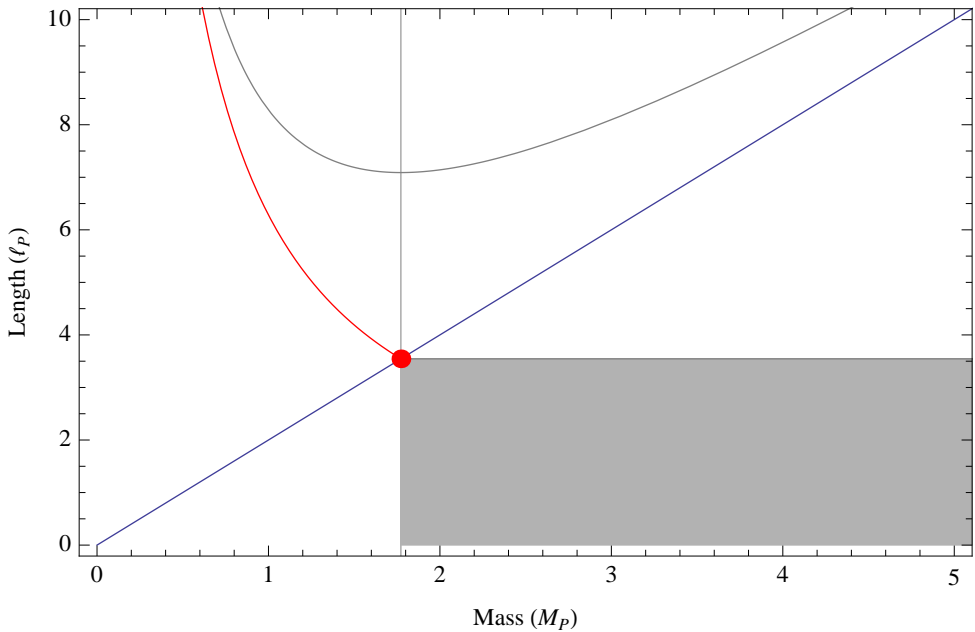}
  \caption{\label{fig:xmBoxGR}Length vs. mass for standard Schwarzschild solution.  The Compton wavelength (red) and horizon radius (blue) curves intersect at $M=\sqrt{\pi} \Mpl,~l=2\sqrt{\pi} \Lpl$ (dot). Equation (\ref{eq:gup1}) approximates the behavior of both these curves (gray). The shaded area is excluded from experiment, while sub-planckian black holes are allowed.}
\end{figure*}

The  essence of self-completeness is also encoded in the generalized uncertainty principle (GUP).  A simple way to  understand the GUP is by considering a light pulse traveling some distance $l$. The  physical measurement of $l$ is affected by an uncertainty $\Delta l_\mathrm{w}\sim \lambda$,  where $\lambda$ is the wavelength of the photon.  

The energy associated with the light pulse can, however, distort the background spacetime.  The measure of $l$ will correspondingly change by an amount
$\Delta l_{\mathrm{g}}\sim  l (|\phi|/c^2)$, 
where $\phi$ is the Newton potential due to a photon of energy $\sim\hbar \nu$, and $c$ is the speed of light. As a result of the above additional uncertainty, one can conclude that the total uncertainty of $l$ is given by
$\Delta l\sim \Delta l_\mathrm{w}+\Delta l_\mathrm{g}\sim \lambda +\Lpl^2/\lambda$.
Such a relation can be derived in several additional \textit{Gedankenexperimente} \cite{maggiore,garay,scardigli,adler,acs01,adler_AJP,casadio} and is corroborated by string theory \cite{string1,string2,string3}. One can additionally extend this line of reasoning to generic particles of mass $M$ to get 
\beq
\Delta x\sim \frac{\hbar}{Mc}+\frac{GM}{c^2} 
\label{eq:gup1}
\eeq
where $\Delta x$ is the position uncertainty (see Fig. \ref{fig:xmBoxGR}).   By minimizing the above expression with respect to the mass, one discovers that the Planck length is again the minimal achievable length scale and that it clearly separates particles (whose size is governed by the Compton wavelength $\sim\hbar/Mc$) from black holes (whose size is governed by the Schwarzschild radius $\sim GM/c^2$). 

\begin{figure*}
\centering
  \includegraphics[scale=1.5]{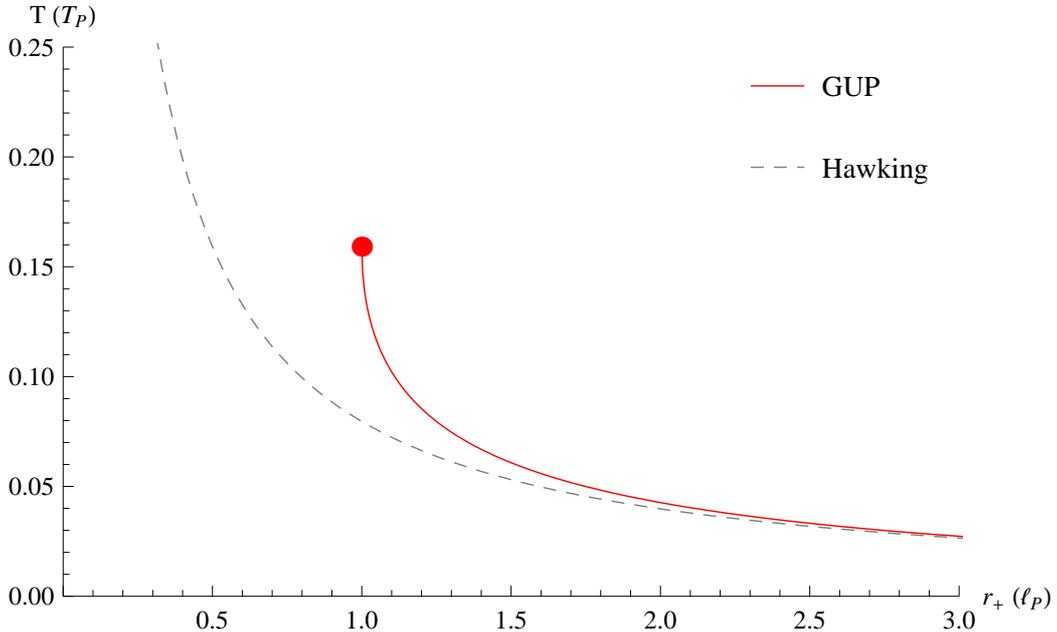}
  \caption{\label{fig:tempOld}Black hole temperature vs. radius in a GUP framework (solid red) eqn. (\ref{eq:gupt}) and Hawking temperature for a regular Schwarzschild black hole (dashed gray). The presence of a hot remnant is indicated by a red dot.}
\end{figure*}

The fact that black holes cannot be smaller than the Planck length and accordingly cannot be lighter than the  Planck mass has repercussions on their emission spectra. The Hawking temperature can be obtained in terms of the energy of the emitted particles as $T\sim Mc^2$.   By assuming in the vicinity of the black hole the uncertainty relation $M\sim \hbar /c\Delta x$ with $\Delta x\sim GM/c^2$, one can readily reproduce the Hawking result.  Taking into consideration the relation (\ref{eq:gup1}), however, the Hawking temperature turns out to be 
\beq
T\sim \frac{\hbar c}{2\pi} \left(\frac{\Delta x}{\Lpl^2}\right) \left(1\pm\sqrt{1-\frac{\Lpl^2}{\Delta x^2}}\right).
\label{eq:gupt}
\eeq
The above equation reproduces the Hawking result  in the limit $\Delta x\gg \Lpl$ if the negative sign is chosen. Equation~(\ref{eq:gupt}) shows relevant modifications when approaching scales $\sim \Lpl$ and implies the existence of  hot Planck scale black hole remnants, as shown in Figure \ref{fig:tempOld} \cite{chen}.

Despite  its virtues, the above analysis is  handicapped by several weak points.  For instance, we implicitly assume that
 quantum gravity effects can be treated semi-classically 
 at scales on the order of the Planck length. On the contrary, one expects that deviations from the classical Schwarzschild radius should occur before the Planck scale, \textit{i.e.} when one reaches energies $\lesssim\Mpl$. This possibility is supported by the inadequacy of the Schwarzschild metric  as an accurate description of the sub-Planckian  spacetime. 
 
 In the particle phase, \textit{i.e.} at energies $<\Mpl$, matter is not sufficiently compressed to collapse into a black hole.  The Schwarzschild metric, however, allows for black holes of any mass and size -- even for $\Mbh<\Mpl$ and $r_{\rm H}<\Lpl$ -- in sharp contrast to the aforementioned self-complete character of gravity. Such limitations of the Schwarzschild metric become more severe  by noting that the temperature (\ref{eq:gupt}) cannot be derived by its surface gravity or from that of any known black hole solution of GR. 
 
The GUP additionally introduces an ambiguity of the sign in Eq.~(\ref{eq:gupt}), whose positive sign choice has no physical meaning. Lastly, the resulting black hole remnants have been conjectured as a natural cold dark matter component.  As mentioned above, these ``remnants'' do not have a vanishing temperature, as one would expect, but a Planckian temperature.    These issues consequently cast doubts about the stability of  such black hole remnants.  By inspecting the heat capacity associated with Eq.~(\ref{eq:gupt}), \textit{i.e.}, $C = dM/dT$, one finds that it is negative and asymptotically vanishes for $r_{\rm H}\to\Lpl$.   This means that the system is suffering from the equivalent instabilities of conventional black hole evaporation. The emission persists as a runaway divergent process up to the Planckian regime.  When $\Mbh\sim T\sim\Mpl$, however, the Schwarzschild metric cannot longer describe the system ``black hole $+$ radiation'' due to relevant quantum back reaction on the metric itself.

 A viable solution to the above problems is offered by those families of quantum gravity improved black hole metrics that admit an extremal configuration even in the neutral, non-rotating case.  Such metrics are inspired by a variety of formulations, including non-commutative geometry (NCG) \cite{nss06,nic09,ns10}, non-local gravity  \cite{mmn11,nicolini1}, asymptotically-safe gravity  \cite{br00}, loop quantum gravity \cite{mod06,mp09,hmp10,carr}, vector ungravity \cite{jrm3} and Bardeen-like, short scale, quantum gravity effects \cite{hay06,eurosmail}. The degeneracy of the horizon allows for a minimum-size extremal black hole and lets one circumvent the above inconsistencies of the Schwarzschild metric. 
 
 As a by-product, the self-complete character of gravity is preserved in the case of black hole decay through Hawking emission.   Contrary to the Schwarzschild metric, in which the curvature singularity can be exposed in the final stage of the evaporation, extremal configurations are zero temperature black holes also stable evaporation remnants. In this spirit, NCG-inspired black holes have been exploited to improve the self-completeness paradigm \cite{euro1}. More recently, a Schwarzschild-like self-complete metric admitting horizon extremisation has been derived solely in the realm of GR without invoking additional principles like NCG, GUP, \textit{etc.} \cite{ns12}. In addition, such a new metric can pave the way to a solution of the recently-uncovered incompatibility between self-completeness and another widely expected character of quantum gravity, namely the spontaneous dimensional reduction of spacetime at the Planck scale \cite{mn13}.    

In this paper, we  further the above line of research and reconcile the ideas of GUP with the self-complete character of gravity in a consistent way.  Rather than considering wavelength corrections as in (\ref{eq:gup1}), we follow the route of implementing a minimal resolution length $\sqrt{\beta}$ at the level of canonical commutators. Taking advantage of the resulting modifications of integration measures in momentum space, we derive a non-local version of the Schwarzschild geometry.  We then exploit the properties of this new metric to draw further conclusions about self-completeness and GUP with special attention to resulting corrections at the Planck scale.


\section{Generalized uncertainty principle} \label{sec:GUP}

In regular quantum mechanics, the cannonical commutator,
\beq \label{eq:hup_com}
\left[\mathbf{x}, \mathbf{p}\right] = i \hbar ~,
\eeq
results in Heisenberg's well-known uncertainty relation between position and momentum
\beq
\Delta x \Delta p \geq \frac{\hbar}{2}~.
\eeq
However, if additional momentum dependent terms are added to Eq. (\ref{eq:hup_com}),
\beq
[\mathbf{x}^i,\mathbf{p}_j] = i \delta^i_j \hbar (1+\beta \mathbf{p}^2)~,
\eeq
($\beta>0$) this will result in a modified uncertainty relation of the form
\beq \label{eq:gup}
\Delta x \Delta p \geq \frac{\hbar}{2} \left( 1 + \beta(\Delta p)^2 \right)~.
\eeq
Such modification is known in the literature as the GUP. In turn, Eq. (\ref{eq:gup}) introduces a non-zero commutation between the coordinate operators
\beq
\left[\mathbf{x}_i, \mathbf{x}_j \right] = 2i \hbar \beta \left(\mathbf{p}_i \mathbf{x}_j - \mathbf{p}_j \mathbf{x}_i \right)~.
\eeq
Because this commutator is non-vanishing unless $\beta=0$, the GUP introduces a non-zero minimal uncertainty in position, which translates into the existence of a minimal length $\sqrt{\beta}$ (for recent reviews on the huge literature in this field see \cite{frankfurt,sabine}). This implies that position eigenstates cannot exist and it is necessary to work with momentum eigenstates or limit ourselves to minimal-uncertainty position states \cite{kempf}. Furthermore, this results in a momentum integration measure
\beq
\int{\frac{d^n p}{1+\beta \vec{p}^2} |p \rangle \langle p|}=1~,
\eeq
which presents a UV cutoff of $\sqrt{\beta}$, where $n$ is the Euclidean space dimension \cite{kempf}. 

GUP approaches have found a myriad of applications in high energy physics and quantum systems, including quantum field theory \cite{crowell,viqar}, gauge theories \cite{kober}, cosmology \cite{zhu,kim} and particle physics \cite{gup1}.  Applications to black hole thermodynamics are of particular interest in the present context, and the interested reader is referred to \cite{scardigli,acs01,gupbh4,gupbh3,gupbh2,gupbh1,gup2,misha1,misha2,misha3} and references therein. 

For our purposes, the implementation of GUP effects in the gravitational field requires certain discussion \cite{nicolini1}. The suppression of the UV sector corresponds to a non-local deformation of the integration measure due to the action of a infinite number of derivative terms.  As a result, GUP deformations can be encoded in non-local gravity actions.  Such actions have been proposed with the goal of formulating a perturbative, super-renormalizable, UV finite  approach to quantum gravity. In \cite{kras,tomb}, the following non-local Lagrangian has been proposed:
\bea
{\cal L}_\mathcal{G}& =& \sqrt{-g}\:\left\{\:\frac{\beta}{\kappa^2}R 
- \beta_2(R_{\mu\nu}R^{\mu\nu} - \frac{1}{3}R^2) + \beta_0R^2 + 
\tilde{\lambda} \right.\nonumber\\
 & &+\left.\left(R_{\mu\nu}\,h_2(-\frac{\tilde{\Box}}{\Lambda^2})
\,R^{\mu\nu} -\frac{1}{3}R\,h_2(-\frac{\tilde{\Box}}{\Lambda^2})\,R
\right) - R\,h_0(-\frac{\tilde{\Box}}{\Lambda^2})\,R
\:\right\}\nonumber\\
 & & -\frac{1}{2\xi}f^\mu[g]w(-\frac{\nabla^2}{\Lambda^2})
f_\mu[g] + \bar{c}^\mu M_{\mu\nu}c^\nu \label{eq:gact},
\eea
where $\tilde{\Box}=\nabla^\mu\nabla_\mu$ and $\nabla^2$  respectively denote the 
covariant and ordinary D'Alembertian,  
$f_\mu[g]$ is the gauge-fixing function with gauge-term weight 
$w$, $\bar{c}^\mu M_{\mu\nu}c^\nu$ is the Faddeev-Popov term, $\kappa^2= 16 \pi G$, $\Lambda$ is some energy scale, $\tilde{\lambda}$ is the cosmological constant and $h_0$, $h_2$ are non-polynomial entire functions. The theory has been recently re-proposed in \cite{mod12} and applied to massive gravity \cite{massive}, the Starobinksi model \cite{bmt13} and to resolve the initial cosmological singularity \cite{cmn13}. A complementary formulation leading to the most general covariant, ghost-free gravitational action has been presented in \cite{lancaster}. 

Gravitation is widely expected to be asymptotically-safe \cite{sw}. This implies that, at the fixed point of the theory, interaction terms turn out to be negligible.  One can, as a result, employ truncated versions of the Lagrangian (\ref{eq:gact}) and derive the corresponding field equations by considering just the effects of the modified propagator \cite{bar1,bar2,bar3,moffat}.  From functional variation of the total action
\beq
S=S_\mathrm{G} + S_\mathrm{M}~,
\eeq
one finds
\beq
{\cal A}^2(\square) \left( R_{\mu\nu} - \frac{1}{2} g_{\mu\nu} R \right) = 8 \pi G T_{\mu\nu}~,
\eeq
where $\mathcal{A}(\square)$ is a non-polynomial entire function (deriving from $h_0$ and $h_2$) of the dimensionless generally covariant D'Alambertian operator, $\square = \ell^2 g_{\mu\nu} \nabla^\mu \nabla^\nu$, with $\ell\equiv 1/\Lambda$.


Following \cite{nicolini1}, the above equations can be cast in a more familiar form as
\beq
\label{nlee}
R_{\mu\nu} - \frac{1}{2} g_{\mu\nu} R = 8 \pi G \mathcal{T}_{\mu\nu}~,
\eeq
with
$\mathcal{T}_{\mu\nu} \equiv {\cal A}^{-2} (\square) T_{\mu\nu}$. In such a form, non-local effects are encoded into a non-standard source term couple to ordinary Einstein gravity.  In the case of a static, spherically symmetric source, the conventional energy-momentum tensor displays an energy density peaked at the origin \cite{jrm3,gaete}, \textit{i.e.},
\beq \label{eq:regT}
T^0_0=-\frac{M}{4 \pi r^2} \delta (r)~,
\eeq
where $\delta(r)$ is the Dirac delta function.  The line element solving (\ref{nlee}) will be static and spherically symmetric as usual:
\beq \label{eq:sphds}
ds^2=-f(r)dt^2 + f^{-1}(r)dr^2 + r^2 d\Omega^2,
\eeq
\beq
f(r) = 1- \frac{2 G{\cal M}(r) }{r}~,
\eeq
with the unknown function $\mathcal{M}(r)$,
\begin{equation}
{\cal M}(r) = - 4\pi \int_0^r dr^\prime r^{\prime 2}\ {\cal T}^0\,_0~,
\label{mass}
\end{equation}
accounting for all non-local effects and necessarily satisfying $\mathcal{M}(r) \rightarrow M$ for $r\gg\ell$, where $M$ is total mass-energy of the system.

In order to find $\mathcal{M}(r)$, it is necessary to choose a particular $\mathcal{A}(\square)$. 
Unfortunately, there is to date no experimental information about quantum gravity and we possess no experimental restrictions on $\mathcal{A}$.  We can nevertheless postulate the profile of the cuf-off function by invoking some reasonable physical principle.  Along this line of reasoning, one can model the effect of the GUP by requiring the action of $\mathcal{A}^{-2}$ on $T^0_0$ to be given by
\beq
\mathcal{A}^{-2}(\square) \delta(\vec{x}) = (2\pi)^{-3} \int \frac{d^3p}{1+\beta \vec{p}^2}e^{i\vec{x} \cdot \vec{p}}~,
\label{eq:Aaction}
\eeq
where $\vec{x}$ are free-falling, Cartesian-like coordinates, provided that $\beta=\ell^2$. From (\ref{eq:Aaction}) it follows that the profile of ${\cal A}$ must be
\beq
{\cal A}(\Box) = (1 -  \Box)^{1/2}~.
\eeq
By means of the Schwinger representation, the exponentiation of a generic differential operator $\hat{\Delta}$ can be written as
\beq 
\hat{\Delta}^{\alpha} = \frac{1}{\Gamma(-\alpha)}\int \limits_0^\infty ds~s^{-\alpha-1} \ e^{-s \hat{\Delta}}~.
\eeq
As a consequence, by setting $\hat{\Delta}=1-\square$ and $\alpha=1/2$, one can represent $\mathcal{A}$ as
\beq 
(1 -  \Box)^{1/2} = -\frac{1}{2\sqrt{\pi}}\int \limits_0^\infty ds~s^{-3/2}\ e^{-s}e^{s\square}.
\eeq
The above expression reconciles the GUP and non-local gravity: it is evident that ${\cal A}$ acts as a non-polynomial entire function. Accordingly, $\mathcal{A}^{-2}$ can be obtained from the case  $\alpha=-1$. 

It is now straightforward to compute the energy density by applying the operator on the standard stress-energy tensor:
\beq \label{eq:modT}
\mathcal{T}^0_0=- M \mathcal{A}^{-2}(\square) \delta(\vec{x}) = -\frac{M}{\beta} \frac{e^{-|\vec{x}|/\sqrt{\beta}}}{4\pi |\vec{x}|}~.
\eeq

\begin{figure*}
\centering
  \includegraphics[scale=1.5]{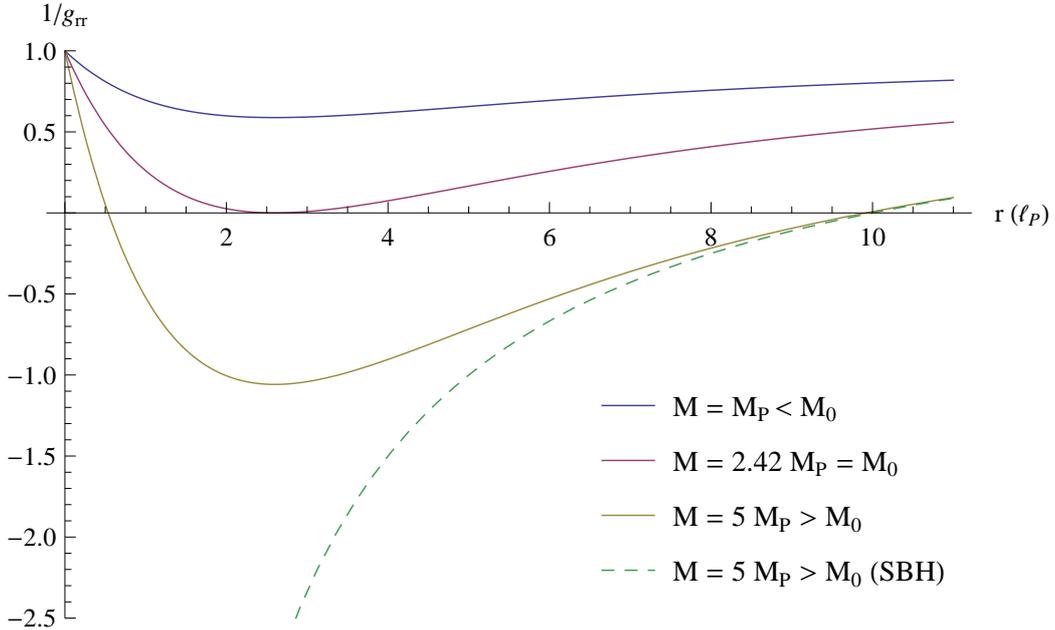}
  \caption{\label{fig:grr}Metric coefficient for GUP metric (\ref{eq:GUP_metric}) with $\sqrt{\beta} = 1.45 \Lpl$. Notice naked singularity, extremal and regular black hole cases. The Schwarzschild (SBH) case for $M=5 \Mpl$ is showed for comparison. The minimum of the extremal case takes place at $M_0\approx 1.67 \sqrt{\beta}c^2/G$ and a $r_0\approx 1.79 \sqrt{\beta}$ for all values of $\beta$.}
\end{figure*}

Finally, integrating (\ref{eq:modT}) we find
\beq
\mathcal{M}(r)/M = 1 - e^{-r/\sqrt{\beta}} -(r/\sqrt{\beta}) e^{-r/\sqrt{\beta}}~,
\eeq
which means, by substitution in (\ref{eq:sphds}), that the GUP inspired metric is given by
\beq \label{eq:GUP_metric}
ds^2=-\left(1-2\frac{GM}{c^2 r}\gamma(2;r/\sqrt{\beta})\right)dt^2-\left(1-2\frac{GM}{c^2 r}\gamma(2;r/\sqrt{\beta})\right)^{-1}dr^2+r^2d\Omega^2
\eeq
where $\gamma(s;x)=\int^x_0 t^{s-1} e^{-t}dt$ is the lower incomplete gamma function. The spacetime (\ref{eq:GUP_metric}) matches the Schwarzschild metric at large distances, ($r\gg \sqrt{\beta}$). However the horizon structure is different.
The corresponding metric coefficient is shown in Fig. \ref{fig:grr}.

By studying the horizon equation $g_{rr}^{-1}=0$ we can distinguish three cases depending on the value of the total mass $M$ with respect to a mass scale $M_0$:
\begin{enumerate}[i)]
\item for $\Mbh=M>M_0$ we have two horizons $r_\pm$. In the limit when $M>>M_0$, the outer radius coincides with the standard value ($r_+ \to 2GM/c^2$),  while the inner one vanishes ($r_-\to 0$);
\item for $\Mbh=M=M_0$ the two horizons coalesce into a single degenerate horizon $r_+=r_-=r_0$, corresponding to an extremal black hole solution;
\item for $M<M_0$ the horizon equation cannot be solved and one has a horizon-less geometry.
\end{enumerate}
Note that the extremal configuration has a mass $M_0\approx 1.66 \sqrt{\beta}c^2/G$ and a radius $r_0\approx 1.73 \sqrt{\beta}$ (Fig. \ref{fig:grr}). 
Finally, at short scale, $r\approx 0$, the curvature singularity is softened but persists. This means that the vacuum energy associated to the virtual graviton exchange is divergent or, in other words, that the graviton propagator is not UV finite. One can verify this by looking at the short scale behavior of the energy density in (\ref{eq:modT}): GUP effects can spread the Dirac into a distribution that is less pathological but still divergent as $r^{-1}$.  We note that an unpleasant drawback of this is the exposure of the (naked) singularity in the horizon-less geometry case ($M<M_0$).

The above results do not come as a surprise. The UV finiteness of any non-local theory like that in (\ref{eq:gact}) is guaranteed at any order only for a certain degree of convergence of the entire function ${\cal A}$. According to the definition given in \cite{efimov1,efimov2}, such a global convergence occurs for entire functions of order higher than $1/2$. As an example, NCG inspired black holes \cite{nss06,nic09,ns10,rizzo,ncrn,ssn09,euroanais10,mn10,jrmpn11} and the associated quantum field theory \cite{ss04,ssn06,kn10} are non-local formulations employing such a kind of entire function  \cite{mmn11}.  At the level of free fields the convergence is achieved also in the case of order $1/2$. However, one can show that the GUP is represented by an entire function (\ref{eq:Aaction}) of order lower than $1/2$, \textit{de facto} failing to improve the classical spacetime geometry \cite{nicolini1}. For a full analysis of the geometry and the thermodynamics of the solution (\ref{eq:GUP_metric}) see \cite{nicolini1}.

Despite the fact that the GUP inspired gravity fails to be UV finite, we wonder whether it may be at least self-complete, \textit{i.e.} whether it is ``always'' able to mask this bad short-distance behaviour behind an event horizon. If this were the case also the previously raised issue of the naked singularity would turn to be circumvented.


\section{Self-completeness} \label{sec:SC}
The metric (\ref{eq:GUP_metric}) is an important step forward \textit{en route} to a reconciliation between GUP and self-completeness. The presence of an extremal configuration naturally prevents the existence of black holes smaller than $r_0$.  Furthermore, in the case of Hawking emission the usual black hole temperature definition $T=\kappa/2\pi$, where $\kappa$ is the surface gravity of the metric (\ref{eq:GUP_metric}), gives
\begin{equation}
T=\frac{\hbar c}{4\pi r_+}\left(1-\frac{r_+^2}{\beta} \frac{ e^{-r_+/\sqrt{\beta} }}{\gamma(2;r/\sqrt{\beta})} \right).
\label{newtemp}
\end{equation}
This temperature improves the result in (\ref{eq:gupt}), which cannot be associated to any surface gravity.  (\ref{newtemp}) possesses a zero for a finite, positive value of $r_+$.  Such a zero implies the existence of an evaporation remnant and has to coincide with the radius $r_0$ of the extremal configuration according to a general property of the horizon extremisation.   This is a first step in the direction of self-completeness: one cannot probe the curvature singularity during the process of black hole decay.

\begin{figure*}
\centering
  \includegraphics[scale=1.5]{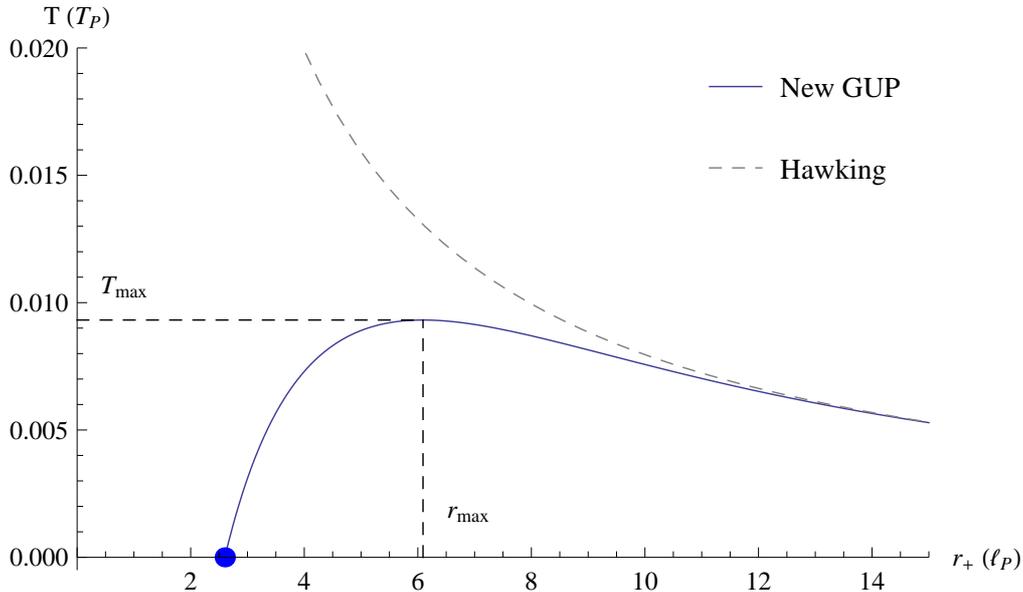}
  \caption{\label{fig:newtemp} New GUP black hole temperature eq. (\ref{newtemp}) for $\sqrt{\beta} =1.45 \Lpl$ (solid blue) and the regular Hawking temperature (dashed gray). The black hole achieves a maximum temparture $T_\mathrm{max}\approx9.34\times10^{-3} T_\mathrm{P}$ at $r_\mathrm{max}\approx4.20\sqrt{\beta}$. Unlike the old GUP temperature (cf. Fig. \ref{fig:tempOld}), our new solution yields a cold remnant (blue dot).}
\end{figure*}

We notice that such an evaporation end-point exhibits intriguing new properties.  At $r_+=r_{\rm max}\approx 4.20 \sqrt{\beta}$ the temperature admits a maximum $T_{\rm max}\equiv T(r_{\rm max})\approx 1.35\times 10^{-2} \hbar c / \sqrt{\beta}$. This fact has important repercussions for the stability of the evaporation remnant.   By examining the form of the heat capacity  
\[
C=\frac{\partial M}{\partial r_+}\left(\frac{\partial T}{\partial r_+}\right)^{-1}
\]
one can distinguish three regimes: $C<0$ for $r_+>r_{\rm max}$,  $C\to\pm\infty$ for $r_+\to (r_{\rm max})_\mp$ and $C>0$ for $r_0<r_+<r_{\rm max}$.
The  profile of $C$ is controlled by the derivative of the temperature (sign and extremal points), being $\partial M/\partial r_+0$ positive and finite for $r_+>r_0$ (see Fig \ref{fig:newtemp}).  From the above analysis one can conclude that, at the maximum temperature $T_{\rm max}$, the system undergoes a  transition from an unstable negative heat capacity phase to a stable positive heat capacity cooling down towards a cold extremal configuration. The latter is characterized by both vanishing temperature and vanishing heat capacity ($\partial M/\partial r_+=0$ for $r_+=r_0$) becoming a reliable candidate for cold dark matter component. We stress that during the process no relevant quantum back reaction occurs and no further short scale corrections have to be taken into account for the metric (\ref{eq:GUP_metric}). This can be seen by noting that $T\ll\Mbh$ during all the evaporation, being $T/\Mbh<T_{\rm max}/M_0 \approx 8.06 \times 10^{-3} G \hbar/ (\beta c)$.

To prove that the above scenario correctly describes the self-complete character of gravity, however, we need to show how the transition ``particle $\leftrightarrow$ black hole'' takes place.

%
%


Following the prescription outlined in  \cite{euro1}, we start by deriving the radius of the extremal configuration. From the horizon condition $1/g_{rr}=0$ one can define the mass parameter $M$ as a function of the radius $r_+$,
\beq
M\equiv\Mbh(r_+)=\frac{c^2}{2G}\frac{r_+}{\gamma(2; r_+/\sqrt{\beta})}.
\label{rplus}
\eeq
The minimum of this function can be calculated by considering $dM(r_+)/dr_+=0$, whose solution $r_0$, given by 
\beq
\gamma(2; r_0/\sqrt{\beta})-\left(\frac{r_0}{\sqrt{\beta}}\right)^2 e^{-r_0/\sqrt{\beta}}=0,
\label{extremalrad}
\eeq
identifies the radius of the extremal configuration for which the temperature (\ref{newtemp}) vanishes as expected.

\begin{figure*}
\centering
  \includegraphics[scale=1.2]{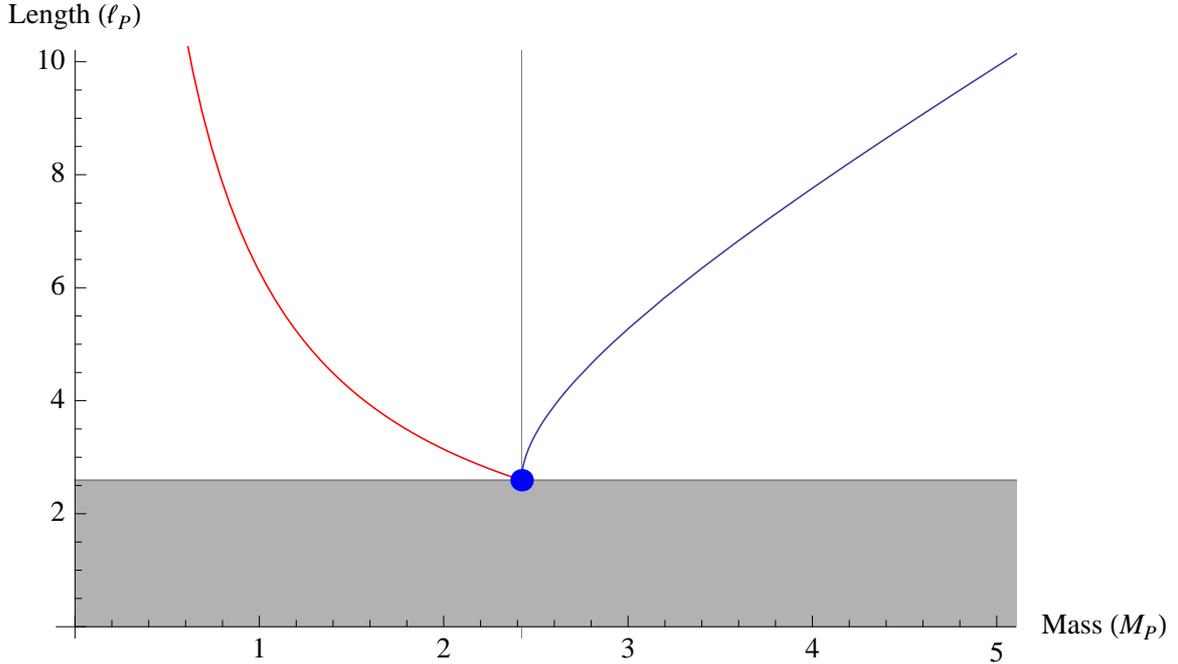}
  \caption{\label{fig:xmBoxGUP}Plot of length vs. mass including new GUP corrections for $\sqrt{\beta}\approx1.45 \Lpl$. The Compton wavelength (red) and horizon radius (blue) curves intersect at $(M_0, r_0)$, marked by a dot. The shaded area is excluded from experiment, meaning there can never be an exposed singularity.}
\end{figure*}

 Black holes can have radii $r_+\geq r_0$, while at shorter scales the horizon equation has no solutions. That is: for $r_+\leq r_0$ only quantum mechanical particles can exist. As a result, we assume that $r_0$ is the transition point between the two aforementioned phases. This fact is summarized in a the condition
\beq
 \frac{h}{cM_0}=r_0.
 \label{gupself}
\eeq
where $M_0\equiv \Mbh(r_0)$.
We note that Eq.~(\ref{extremalrad}) is independent of the parameter $\beta$ and can be solved in terms of the dimensionless quantity $x_0\equiv r_0/\sqrt{\beta}$. This allows us to fix the value of the parameter $\beta$ in order to fulfil Eq. (\ref{gupself}) as 
\beq
\beta=4\pi\ \frac{\gamma(2;x_0)}{x_0^2} \ \Lpl^2.
\eeq
By introducing the dimensionless quantity $m_0\equiv M_0G/\sqrt{\beta}c^2$, one can write the above relation as $\beta = (2\pi/x_0m_0)\Lpl^2$. Accordingly we obtain
\beq
r_0=\sqrt{\frac{2\pi x_0}{m_0}}\ \Lpl \quad M_0=\sqrt{\frac{2\pi m_0}{x_0}}\ \Mpl. 
\eeq
Recalling that numerical estimates give $x_0\approx 1.79$ and $m_0\approx 1.68$, we obtain $\sqrt{\beta}\approx 1.45\Lpl$, $r_0\approx 2.59\Lpl$ and $M_0\approx 2.42\Mpl$ (see Figure \ref{fig:xmBoxGUP}).


From here on, we can promote $r_0$ and $M_0$  as the new ``fundamental scales''.  Indeed, these parameters identify a consistent transition between the two phases in both directions, \textit{i.e.} during the compression (``particle $\rightarrow$ black hole '') and during the decay (``particle $\leftarrow$ black hole''). We stress that the decay is correctly described in terms of thermal emission at the temperature associated with the surface gravity of the metric (\ref{eq:GUP_metric}) without any ambiguity.  In addition, the singularity can never be exposed during any of the two aforementioned processes, a fact that virtually  eliminates the threat of a of naked singularity for $M<M_0$.

\subsection{Wavelength correction}

 \begin{figure*}
\centering
  \includegraphics[scale=1.1]{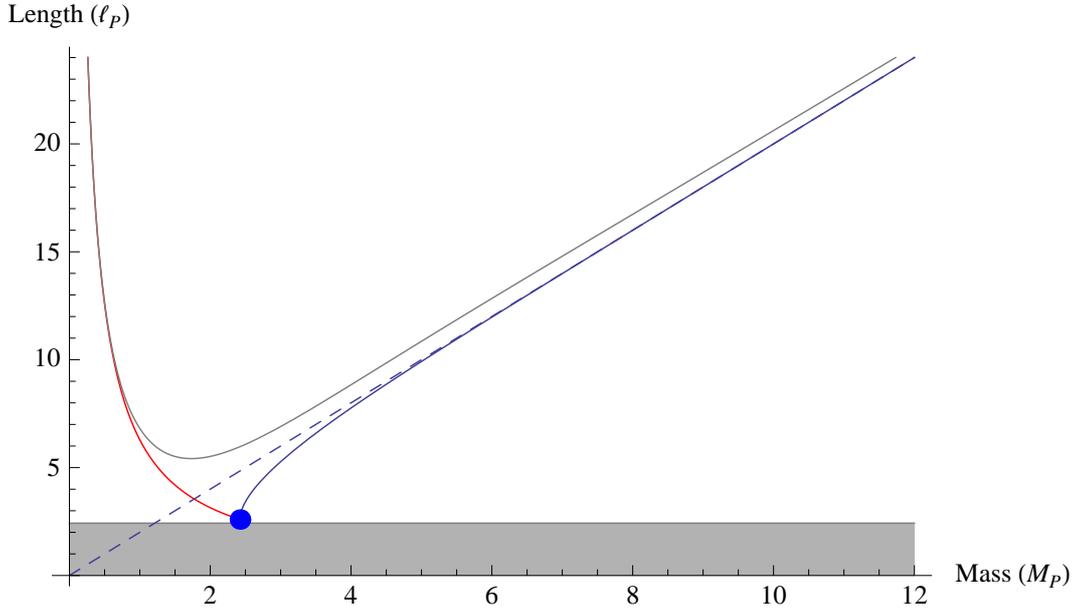}
  \caption{\label{fig:xm} Length vs. mass plot including GUP corrections for $\sqrt{\beta}\approx1.45 \Lpl$. Relation (\ref{gupimproved}), shown in gray, approximates the behavior of both the Compton wavelength (red) and the GUP horizon radius (blue). The shaded area is excluded from experiment, meaning there can never be an exposed singularity.}
\end{figure*}

In light of the above results, we are now ready to re-formulate the \textit{Gedankenexperiment} described in the introductory section.   By writing Eq.~(\ref{eq:GUP_metric}) as
\beq
ds^2=-(1+2\phi_{\mathrm{GUP}})dt^2+(1+2\phi_{\mathrm{GUP}})^{-1}dr^2+r^2d\Omega^2
\eeq

one obtains an improved Newtonian potential $\phi_{\mathrm{GUP}}$ that linearly vanishes at the origin, 
$\phi_{\mathrm{GUP}}\sim -GMr/\beta$,  and matches the standard Newtonian potential, $\phi_{\mathrm{GUP}}\approx -GM/r$ at large distances.  Such a quantity allows us to estimate the local spacetime distortion in terms of the GUP inspired non-local gravity, rather than in terms of standard Einstein gravity.  As a result, one obtains a gravitational uncertainty $\Delta \lambda_\mathrm{g}=2\pi\Lpl^2 \gamma(2,2\pi\hbar G/c\lambda\sqrt{\beta})/\lambda$. By considering the full uncertainty for an arbitrary massive particle, one can write
\beq
\Delta x\sim 2\pi\frac{\hbar}{Mc}+2\frac{GM}{c^2} \gamma\left(2;\Delta x/
\sqrt{\beta}\right)
\label{gupimproved}
\eeq

 in place of (\ref{eq:gup1}). As shown in Fig. \ref{fig:xm}, away from the Planck scale, the above relation works as (\ref{eq:gup1}), namely $\Delta x\approx 2\pi\frac{\hbar}{Mc}$ for quantum particles ($M\ll\Mpl$), and $\Delta x\approx \frac{GM}{c^2}$  for classical black holes ($M\gg\Mpl$).  At the Planck scale ($M\sim\Mpl$), however, the gamma function in (\ref{gupimproved}) departs from unity, $0<\gamma\left(2;\Delta x(\Mpl)/
\sqrt{\beta}\right)<1$. This corresponds to accounting 
for a crucial non-local gravity effect, namely the minimal black hole mass $M_0$.  One then finds that for $M\sim M_0$
\beq
\Delta x\approx 2\pi\frac{\hbar}{M_0c}+2\frac{GM_0}{c^2} \gamma\left(2;r_0/
\sqrt{\beta}\right)=2r_0
\eeq
We stress that, contrary to the case in (\ref{eq:gup1}), the scale $M_0$ is corroborated by the corresponding metric. In this sense (\ref{gupimproved}) provides a Planck scale completion of (\ref{eq:gup1}).
 

As a related comment we note that  (\ref{gupimproved}) is not in conflict with the uncertainty relations in (\ref{eq:gup}). Rather, it is the ``translation'' of the deformed integration measure in (\ref{eq:Aaction}) from locally flat coordinates to curvilinear ones. In such a transformation, the GUP inspired non-local gravity works in a more complicated way than Einstein gravity, by introducing nontrivial terms like the incomplete gamma function.



\section{Conclusions} \label{sec:conclusions}
In this paper we showed how to reconcile the self-complete character of gravity with the GUP.   We started by stressing that the conventional ideas at the heart of the GUP fail to be accurate at the Planck scale.  Among these various limitations, the GUP implies the existence of black hole remnants that are not compatible with a neutral, classical metric like the Schwarzschild geometry.  As a result, one ends up with an ambiguity between particles and black holes in the sub-Planckian regime.

Against this background, we exploited the idea of GUP at the level of integration measure in momentum space in order to construct a non-local version of Einstein's equations.  By deriving the corresponding static, neutral black hole solution, we showed that Planck scale black hole remnants naturally emerge from the metric coefficients as extremal zero temperature configurations. This fact paves the way to a consistent scenario for the self-completeness that overcomes the standard case limitations.   Black holes form as a result of matter compression to sizes of the order of the radius of the extremal configuration ($\sim\Lpl$). A further increase of energy leads to bigger black holes that approach classical solutions of GR.  

The reverse process is also free from pathologies.  A black hole cannot endlessly decay.  The evaporation end-point is represented again in terms of the aforementioned extremal configuration, which fulfils the special and unique feature of being at the same time the heaviest quantum particle and the lightest black hole.  In addition, they enjoy the property of having both zero temperature and zero heat capacity, thus becoming a reliable candidate for dark matter component.

In principle GUP deformations of the integration measure in momentum space could be exploited to account for further corrections to the spectra of particles emitted by the black hole.   Preliminary studies in this direction concerning the case of NCG-inspired black holes, however,  show that these kind of corrections lead only to sub-leading effects \cite{pieroeli}. Such a result is consistent with the general property of metrics admitting a
maximum black hole temperature: the nature of the radiation is of secondary concern being the quantum backreaction negligible during the complete evaporation process.

Lastly, we considered a \textit{Gedenkenexperiment} that summarizes the above results and improves the conventional reasoning.   We introduced a new GUP that improves the conventional relations presented in \cite{maggiore,garay,scardigli,adler,acs01,adler_AJP} with non-local gravity corrections at the Planck scale.   This satisfies all limiting cases for the expected black hole behavior by replacing standard Einstein gravity with the GUP inspired version.

%

\subsection*{Acknowledgments}MI and JM would like to thank the generous hospitality of the Frankfurt Institute for Advanced Studies, at which this work was initiated. This work has been supported by the project ``Evaporation of microscopic black holes'' (PN) of the German Research Foundation (DFG), by the Helmholtz International Center for FAIR within the framework of the LOEWE program (Landesoffensive zur Entwicklung Wissenschaftlich-\"{O}konomischer Exzellenz) launched by the State of Hesse (PN), partially by the European Cooperation in Science and Technology (COST) action MP0905 ``Black Holes in a Violent Universe'' (PN). The authors thank Bernard Carr and Leonardo Modesto for valuable discussions about the manuscript.



\begin{thebibliography}{99}

\bibitem{kempf} A.~Kempf, G.~Mangano and R.~B.~Mann, ``Hilbert space representation of the minimal length uncertainty relation,''
  Phys.\ Rev.\ D {\bf 52}, 1108 (1995)
  [hep-th/9412167].

\bibitem{dvali1}  G.~Dvali and C.~Gomez,``Self-Completeness of Einstein Gravity,''
  arXiv:1005.3497 [hep-th].
  
\bibitem{dvali2}  G.~Dvali, S.~Folkerts and C.~Germani, ``Physics of Trans-Planckian Gravity,''
  Phys.\ Rev.\ D {\bf 84}, 024039 (2011)
  [arXiv:1006.0984 [hep-th]].

\bibitem{dvali3}  G.~Dvali and C.~Gomez, ``Ultra-High Energy Probes of Classicalization,''
  JCAP {\bf 1207}, 015 (2012)
  [arXiv:1205.2540 [hep-ph]].
  
\bibitem{euroantonio} 
  A.~Aurilia and E.~Spallucci,
  ``Planck's uncertainty principle and the saturation of Lorentz boosts by Planckian black holes,''
  arXiv:1309.7186 [gr-qc].

\bibitem{maggiore} 
  M.~Maggiore,
  ``A Generalized uncertainty principle in quantum gravity,''
  Phys.\ Lett.\ B {\bf 304}, 65 (1993)
  [hep-th/9301067].
  
\bibitem{garay} 
  L.~J.~Garay,
  ``Quantum gravity and minimum length,''
  Int.\ J.\ Mod.\ Phys.\ A {\bf 10}, 145 (1995)
  [gr-qc/9403008].


\bibitem{scardigli}  
F.~Scardigli,
  ``Generalized uncertainty principle in quantum gravity from micro - black hole Gedanken experiment,''
  Phys.\ Lett.\ B {\bf 452}, 39 (1999)
  [hep-th/9904025].

\bibitem{adler} 
  R.~J.~Adler and D.~I.~Santiago,
  ``On gravity and the uncertainty principle,''
  Mod.\ Phys.\ Lett.\ A {\bf 14}, 1371 (1999)
  [gr-qc/9904026].

\bibitem{acs01} 
  R.~J.~Adler, P.~Chen and D.~I.~Santiago,
  ``The Generalized uncertainty principle and black hole remnants,''
  Gen.\ Rel.\ Grav.\  {\bf 33}, 2101 (2001)
  [gr-qc/0106080].

\bibitem{adler_AJP} R.~J.~Adler, ``Six easy roads to the Planck scale,''
  Am.\ J.\ Phys.\  {\bf 78}, 925 (2010)
  [arXiv:1001.1205 [gr-qc]].
  
\bibitem{casadio} 
  R.~Casadio and F.~Scardigli,
  ``Horizon wave-function for single localized particles: GUP and quantum black hole decay,''
  arXiv:1306.5298 [gr-qc].

  
\bibitem{string1} 
  G.~Veneziano,
  ``A Stringy Nature Needs Just Two Constants,''
  Europhys.\ Lett.\  {\bf 2}, 199 (1986).

\bibitem{string2} 
  D.~Amati, M.~Ciafaloni and G.~Veneziano,
  ``Can Space-Time Be Probed Below the String Size?,''
  Phys.\ Lett.\ B {\bf 216}, 41 (1989).
  
\bibitem{string3} 
  A.~Aurilia and E.~Spallucci,
  ``Why the length of a quantum string cannot be Lorentz contracted,''
  arXiv:1309.7741 [hep-th].


\bibitem{chen} 
  P.~Chen and R.~J.~Adler,
  ``Black hole remnants and dark matter,''
  Nucl.\ Phys.\ Proc.\ Suppl.\  {\bf 124}, 103 (2003)
  [gr-qc/0205106].


\bibitem{nss06} 
  P.~Nicolini, A.~Smailagic and E.~Spallucci,
  ``Noncommutative geometry inspired Schwarzschild black hole,''
  Phys.\ Lett.\ B {\bf 632}, 547 (2006)
  [gr-qc/0510112].

\bibitem{nic09} 
  P.~Nicolini,
  ``Noncommutative Black Holes, The Final Appeal To Quantum Gravity: A Review,''
  Int.\ J.\ Mod.\ Phys.\ A {\bf 24}, 1229 (2009)
  [arXiv:0807.1939 [hep-th]].

\bibitem{ns10}  
 P.~Nicolini and E.~Spallucci,
  ``Noncommutative geometry inspired wormholes and dirty black holes,''
  Class.\ Quant.\ Grav.\  {\bf 27}, 015010 (2010)
  [arXiv:0902.4654 [gr-qc]].  
  
 \bibitem{mmn11} 
  L.~Modesto, J.~W.~Moffat and P.~Nicolini,
  ``Black holes in an ultraviolet complete quantum gravity,''
  Phys.\ Lett.\ B {\bf 695}, 397 (2011)
  [arXiv:1010.0680 [gr-qc]].

  \bibitem{nicolini1} P. Nicolini, ``Nonlocal and generalized uncertainty principle black holes,'' [arXiv:1202.2102 [hep-th]]. 
  
    
  \bibitem{br00} 
  A.~Bonanno and M.~Reuter, ``Renormalization group improved black hole space-times,''
  Phys.\ Rev.\ D {\bf 62}, 043008 (2000)
  [hep-th/0002196].
  
  \bibitem{mod06} 
  L.~Modesto,
  ``Loop quantum black hole,''
  Class.\ Quant.\ Grav.\  {\bf 23}, 5587 (2006)
  [gr-qc/0509078].

  
  \bibitem{mp09} 
  L.~Modesto and I.~Premont-Schwarz,
  ``Self-dual Black Holes in LQG: Theory and Phenomenology,''
  Phys.\ Rev.\ D {\bf 80}, 064041 (2009)
  [arXiv:0905.3170 [hep-th]].
  
  \bibitem{hmp10} 
  S.~Hossenfelder, L.~Modesto and I.~Premont-Schwarz, ``A Model for non-singular black hole collapse and evaporation,''
  Phys.\ Rev.\ D {\bf 81}, 044036 (2010)
  [arXiv:0912.1823 [gr-qc]].
  
  \bibitem{carr} B.~Carr, L.~Modesto and I.~Premont-Schwarz,
  ``Generalized Uncertainty Principle and Self-dual Black Holes,''
  arXiv:1107.0708 [gr-qc].

  
  \bibitem{jrm3} J.~R.~Mureika and E.~Spallucci, ``Vector unparticle enhanced black holes: exact solutions and thermodynamics,''
  Phys.\ Lett.\ B {\bf 693}, 129 (2010)
  [arXiv:1006.4556 [hep-ph]].
  
  \bibitem{hay06} 
  S.~A.~Hayward, ``Formation and evaporation of regular black holes,''
  Phys.\ Rev.\ Lett.\  {\bf 96}, 031103 (2006)
  [gr-qc/0506126].


  \bibitem{eurosmail} E.~Spallucci and A.~Smailagic,
  ``Black holes production in self-complete quantum gravity,''
  Phys.\ Lett.\ B {\bf 709}, 266 (2012)
  [arXiv:1202.1686 [hep-th]].
  


\bibitem{euro1} E.~Spallucci and S.~Ansoldi,
  ``Regular black holes in UV self-complete quantum gravity,''
  Phys.\ Lett.\ B {\bf 701}, 471 (2011)
  [arXiv:1101.2760 [hep-th]].
  
\bibitem{ns12} 
  P.~Nicolini and E.~Spallucci,
  ``Holographic screens in ultraviolet self-complete quantum gravity,''
  arXiv:1210.0015 [hep-th].
  
  \bibitem{mn13} 
  J.~Mureika and P.~Nicolini,
  ``Self-completeness and spontaneous dimensional reduction,''
  Eur.\  Phys.\  J.\ Plus {\bf 128},  78 (2013)
  [arXiv:1206.4696 [hep-th]].
  
  \bibitem{frankfurt} 
  M.~Sprenger, P.~Nicolini and M.~Bleicher,
  ``Physics on Smallest Scales - An Introduction to Minimal Length Phenomenology,''
  Eur.\ J.\ Phys.\  {\bf 33}, 853 (2012)
  [arXiv:1202.1500 [physics.ed-ph]].
  
  \bibitem{sabine}
  S.~Hossenfelder,
  ``Minimal Length Scale Scenarios for Quantum Gravity,''
   Living Rev. Relativity \textbf{16}, 2 (2013) 
  [arXiv:1203.6191].

\bibitem{crowell}  L.~B.~Crowell,
  ``Generalized uncertainty principle for quantum fields in curved space-time,''
  Found.\ Phys.\ Lett.\  {\bf 12}, 585 (1999).

\bibitem{viqar} V.~Husain, D.~Kothawala and S.~S.~Seahra,
  ``Generalized uncertainty principles and quantum field theory,''
  Phys.\ Rev.\ D {\bf 87}, 025014 (2013)
  [arXiv:1208.5761 [hep-th]].
  
\bibitem{kober} M.~Kober, ``Gauge Theories under Incorporation of a Generalized Uncertainty Principle,''   Phys.\ Rev.\ D {\bf 82}, 085017 (2010)
  [arXiv:1008.0154 [physics.gen-ph]].

\bibitem{zhu}  T.~Zhu, J.~-R.~Ren and M.~-F.~Li,
  ``Influence of Generalized and Extended Uncertainty Principle on Thermodynamics of FRW universe,''   Phys.\ Lett.\ B {\bf 674}, 204 (2009)
  [arXiv:0811.0212 [hep-th]].
  

\bibitem{kim}  W.~Kim, Y.~-J.~Park and M.~Yoon,
  ``Entropy of the FRW universe based on the generalized uncertainty principle,''
  Mod.\ Phys.\ Lett.\ A {\bf 25}, 1267 (2010)
  [arXiv:1003.3287 [gr-qc]].

\bibitem{gup1} 
  S.~Hossenfelder, M.~Bleicher, S.~Hofmann, J.~Ruppert, S.~Scherer and H.~Stoecker,
  ``Collider signatures in the Planck regime,''
  Phys.\ Lett.\ B {\bf 575}, 85 (2003)
  [hep-th/0305262].
  


\bibitem{gupbh4} 
 G.~Amelino-Camelia, M.~Arzano, Y.~Ling and G.~Mandanici,
  ``Black-hole thermodynamics with modified dispersion relations and generalized uncertainty principles,''
  Class.\ Quant.\ Grav.\  {\bf 23}, 2585 (2006)
  [gr-qc/0506110].

\bibitem{gupbh3} Y.~S.~Myung, Y.~-W.~Kim and Y.~-J.~Park,
  ``Black hole thermodynamics with generalized uncertainty principle,''
  Phys.\ Lett.\ B {\bf 645}, 393 (2007)
  [gr-qc/0609031].

\bibitem{gupbh2} A.~Bina, S.~Jalalzadeh and A.~Moslehi,
  ``Quantum Black Hole in the Generalized Uncertainty Principle Framework,''
  Phys.\ Rev.\ D {\bf 81}, 023528 (2010)
  [arXiv:1001.0861 [gr-qc]].
  
\bibitem{gupbh1} S.~Gangopadhyay, A.~Dutta, A.~Saha, ``Generalized uncertainty principle and black hole thermodynamics'' [arXiv:1307.7045 [gr-qc]].


  
  \bibitem{gup2} 
  B.~Koch, M.~Bleicher and S.~Hossenfelder,
  ``Black hole remnants at the LHC,''
  JHEP {\bf 0510}, 053 (2005)
  [hep-ph/0507138].
  
  \bibitem{misha1} 
  M.~Maziashvili,
  ``Black hole remnants due to GUP or quantum gravity?,''
  Phys.\ Lett.\ B {\bf 635}, 232 (2006)
  [gr-qc/0511054].
  
  \bibitem{misha2} 
  D.~Mania and M.~Maziashvili,
  ``Corrections to the black body radiation due to minimum-length deformed quantum mechanics,''
  Phys.\ Lett.\ B {\bf 705}, 521 (2011)
  [arXiv:0911.1197 [hep-th]].

\bibitem{misha3} 
  A.~R.~P.~Dirkes, M.~Maziashvili and Z.~K.~Silagadze,
  ``Black hole remnants due to Planck-length deformed QFT,''
  arXiv:1309.7427 [gr-qc].


\bibitem{kras} 
  N.~V.~Krasnikov,
  ``Nonlocal Gauge Theories,''
  Theor.\ Math.\ Phys.\  {\bf 73}, 1184 (1987)
  [Teor.\ Mat.\ Fiz.\  {\bf 73}, 235 (1987)].
  
  \bibitem{tomb} 
  E.~T.~Tomboulis,
  ``Superrenormalizable gauge and gravitational theories,''
  hep-th/9702146.
  
   \bibitem{mod12} 
  L.~Modesto,
  ``Super-renormalizable Quantum Gravity,''
  Phys.\ Rev.\ D {\bf 86}, 044005 (2012)
  [arXiv:1107.2403 [hep-th]].
 
  
  \bibitem{massive}
  L.~Modesto and S.~Tsujikawa,
  ``Non-local massive gravity,''
  arXiv:1307.6968 [hep-th].
  
  \bibitem{bmt13}
  F.~Briscese, L.~Modesto and S.~Tsujikawa,
  ``Super-renormalizable or finite completion of the Starobinsky theory,''
  arXiv:1308.1413 [hep-th].
  
  \bibitem{cmn13}
  G.~Calcagni, L.~Modesto and P.~Nicolini,
  ``Super-accelerating bouncing cosmology in asymptotically-free non-local gravity,''
  arXiv:1306.5332 [gr-qc].

 
  \bibitem{lancaster} 
  T.~Biswas, E.~Gerwick, T.~Koivisto and A.~Mazumdar,
  ``Towards singularity and ghost free theories of gravity,''
  Phys.\ Rev.\ Lett.\  {\bf 108}, 031101 (2012)
  [arXiv:1110.5249 [gr-qc]].
  
  \bibitem{bar1} 
  A.~O.~Barvinsky,
  ``Nonlocal action for long distance modifications of gravity theory,''
  Phys.\ Lett.\ B {\bf 572}, 109 (2003)
  [hep-th/0304229].
 
\bibitem{bar2} 
  A.~O.~Barvinsky,
  ``On covariant long-distance modifications of Einstein theory and strong coupling problem,''
  Phys.\ Rev.\ D {\bf 71}, 084007 (2005)
  [hep-th/0501093].
 
 \bibitem{bar3} 
  A.~O.~Barvinsky,
  ``Dark energy and dark matter from nonlocal ghost-free gravity theory,''
  Phys.\ Lett.\ B {\bf 710}, 12 (2012)
  [arXiv:1107.1463 [hep-th]].
  

 
\bibitem{moffat}J.~W.~Moffat,
  ``Ultraviolet Complete Quantum Gravity,''
  Eur.\ Phys.\ J.\ Plus {\bf 126}, 43 (2011)
  [arXiv:1008.2482 [gr-qc]].
  
\bibitem{sw} 
  S.~Weinberg,
  ``Ultraviolet Divergences In Quantum Theories Of Gravitation,'' In \textit{General Relativity: An Einstein centenary survey}, ed. S. W. Hawking and W. Israel. Cambridge University Press. pp. 790-831.

\bibitem{gaete} 
  P.~Gaete, J.~A.~Helayel-Neto and E.~Spallucci,
  ``Un-graviton corrections to the Schwarzschild black hole,''
  Phys.\ Lett.\ B {\bf 693}, 155 (2010)
  [arXiv:1005.0234 [hep-ph]].


\bibitem{efimov1}
  G.~V.~Efimov,
  ``Non-local quantum theory of the scalar field,''
  Comm. Math. Phys.\  {\bf 5}, 42 (1967).
  

\bibitem{efimov2}
  G.~V.~Efimov,
  ``On a class of relativistic invariant distributions,''
   Comm. Math. Phys.\  {\bf 7}, 138 (1968).
   
\bibitem{rizzo} 
  T.~G.~Rizzo,
  ``Noncommutative Inspired Black Holes in Extra Dimensions,''
  JHEP {\bf 0609}, 021 (2006)
  [hep-ph/0606051].
  
\bibitem{ncrn} 
  S.~Ansoldi, P.~Nicolini, A.~Smailagic and E.~Spallucci,
  ``Noncommutative geometry inspired charged black holes,''
  Phys.\ Lett.\ B {\bf 645}, 261 (2007)
  [gr-qc/0612035].

   
\bibitem{ssn09} 
  E.~Spallucci, A.~Smailagic and P.~Nicolini,
  ``Non-commutative geometry inspired higher-dimensional charged black holes,''
  Phys.\ Lett.\ B {\bf 670}, 449 (2009)
  [arXiv:0801.3519 [hep-th]].
  
\bibitem{euroanais10} 
  A.~Smailagic and E.~Spallucci,
  ``'Kerrr' black hole: the Lord of the String,''
  Phys.\ Lett.\ B {\bf 688}, 82 (2010)
  [arXiv:1003.3918 [hep-th]].
  
\bibitem{mn10} 
  L.~Modesto and P.~Nicolini,
  ``Charged rotating noncommutative black holes,''
  Phys.\ Rev.\ D {\bf 82}, 104035 (2010)
  [arXiv:1005.5605 [gr-qc]].
  
\bibitem{jrmpn11} 
  J.~R.~Mureika and P.~Nicolini,
  ``Aspects of noncommutative (1+1)-dimensional black holes,''
  Phys.\ Rev.\ D {\bf 84}, 044020 (2011)
  [arXiv:1104.4120 [gr-qc]].

\bibitem{ss04} 
  A.~Smailagic and E.~Spallucci,
  ``Lorentz invariance, unitarity in UV-finite of QFT on noncommutative spacetime,''
  J.\ Phys.\ A {\bf 37}, 1 (2004)
  [Erratum-ibid.\ A {\bf 37}, 7169 (2004)]
  [hep-th/0406174].

\bibitem{ssn06} 
  E.~Spallucci, A.~Smailagic and P.~Nicolini,
  ``Trace Anomaly in Quantum Spacetime Manifold,''
  Phys.\ Rev.\ D {\bf 73}, 084004 (2006)
  [hep-th/0604094].
  
\bibitem{kn10} 
  M.~Kober and P.~Nicolini,
  ``Minimal Scales from an Extended Hilbert Space,''
  Class.\ Quant.\ Grav.\  {\bf 27}, 245024 (2010)
  [arXiv:1005.3293 [hep-th]].


\bibitem{pieroeli} 
  P.~Nicolini and E.~Winstanley,
  ``Hawking emission from quantum gravity black holes,''
  JHEP {\bf 1111}, 075 (2011)
  [arXiv:1108.4419 [hep-ph]].

  
%
%
%
%
%
%
%







  









\end{thebibliography}

\nocite{*}
\end{document}